# A copper negative index metamaterial in the visible / near-infrared

J. Parsons and A. Polman

*Centre for Nanophotonics, FOM Institute AMOLF,*

*Science Park 104, Amsterdam, 1098 XG, The Netherlands*

**ABSTRACT**

We propose a design for a plasmonic copper metamaterial with a negative index of refraction at visible/near-infrared wavelengths. Using numerical simulations, we demonstrate negative refraction by a copper metamaterial prism, and perform a parameter extraction technique to verify the sign of the effective, electric permittivity and magnetic permeability. Our proposed design has a figure of merit comparable to similar silver-based metamaterials operating in the visible/near-infrared range. These findings have implications for the design of low cost plasmonic devices and negative-index metamaterials in the visible / near-infrared.

'Metamaterials' are artificial materials structured on sub-wavelength scales which provide a means to achieve unconventional electromagnetic responses, such as a negative index of refraction. The most general case of a "negative-index" metamaterial requires the simultaneous existence of a negative real (electric) permittivity $\varepsilon$ and negative real (magnetic) permeability $\mu$. Metals are naturally occurring materials which have a negative electric permittivity and interact strongly with an applied electric field, however for most wavelengths there are no materials which exhibit such a response to an applied magnetic field. Instead, a number of structural elements are used to engineer the desired magnetic response. These geometries include split-ring resonators, cut-wire pairs and meshed structures such as the 'fishnet', which have inspired the realisation of negative-index metamaterials at microwave [1], near-infrared [2-4] and visible wavelengths [5-6]. At visible wavelengths, metamaterials are often constructed using silver (Ag) or gold (Au) since these materials possess both large $-\text{Re}(\varepsilon)$ and have losses which are amongst the smallest of all the metallic elements. Despite this, there remains a considerable interest to utilise alternatives such as alkali metals or doped semiconductors to further minimise losses, decrease the operating wavelength, or reduce fabrication costs [7-8]. Recently, it has also been suggested that copper (Cu) may be suitable for building low cost plasmonic structures, since Cu nanoparticles support localised surface plasmon resonances (LSPRs) in the near-infrared with quality factors comparable to those of Ag [9].

Here, we propose a Cu negative-index metamaterial in which the magnetic resonance is geometrically tuned to exploit the plasmonic nature of Cu at $\lambda$ = 650 - 750 nm. We consider a multilayer structure wih thickness 240 nm formed from six unit-cells of 20 nm Cu / 20 nm dielectric ($n$ = 1.4 + 0i). The multilayer stack is perforated with a square array of cylindrical holes (array period 300 nm, hole diameter 200 nm) which are filled with vacuum (Figs. 1a,b). Simulations are performed using commercial finite-difference-time-domain software (Lumerical version 6.5) assuming optical constants fitted to reference data for Cu [10].

To study refraction by the metamaterial, a prism is simulated in vacuum (prism angle 0.9°, Fig. 1c). The prism has a 5 μm x 5 μm square base and is tapered in thickness from seven (280 nm) to five (200 nm) unit-cells. In designing the metamaterial, the number of periods were chosen to ensure convergence of the negative index response. Additional simulations (not shown) were initially performed which studied the evolution of effective parameters whilst systematically increased the thickness of the metamaterial. The flat base of the prism is illuminated at normal incidence with a Gaussian beam (diameter 4 μm) aligned through the centre axis, and the appropriate electromagnetic fields are recorded at positions on two planes above and below the prism which are lying perpendicular to the incident wavevector. The directions of the incident electric field and wave vectors are indicated by the dashed arrows.

Fig. 2 shows the simulated transmittance of the planar and prism metamaterial geometries illustrated in Figs. 1b,c. There are a number of resonant features in the spectral transmittance which arise from coupling between LSPRs [11-12]. A single Cu layer perforated with an array of holes supports LSPRs in the metallic regions *between* holes. When two such metallic nanostructures are brought to within the near-field decay length of the LSPR (≈ 10s of nm), the modes of adjacent layers couple such that the composite structure supports a hybridisation of symmetric and asymmetric coupled modes [12-13]. In a bilayer system, the asymmetric mode is red-shifted to wavelengths above the unperturbed LSPR whilst the symmetric mode is blue-shifted below it. The difference between resonant wavelengths is explained by the relative alignments of the dipole moments in adjacent layers which modify the restoring force associated with the dipolar resonance. Furthermore, coupling to the asymmetric mode is achieved using the incident magnetic field, since the change in phase of the electric field over the distance separating the two layers is less than the π phase change required to excite this mode. For the twelve-layer system considered here, the bilayer hybridisation principle also applies, with the exception that there are additional permutations of symmetric / asymmetric coupling, leading to the formation of a transmission band.

The resonant excitation of the asymmetric LSPR through the incident magnetic field gives rise to an effective magnetic permeability at visible wavelengths. Since Cu naturally possesses a negative electric permittivity, it is reasonable to expect negative index behaviour. To test this hypothesis one can simulate the refraction of light impinging on the flat side of a prism. The angular distribution of the transmitted light is obtained by performing a near-to-far-field transformation of the time-averaged field distributions at output side of the prism. Fig. 3 shows the far-field electric field intensity as a function of incident wavelength $\lambda$ and refraction angle $\theta_{ref}$ with respect to the surface normal. For clarity, the far-field intensities have been normalised by the appropriate transmittance curve in Fig. 2. Note that for $\lambda = 450$ nm, refraction is positive. The refraction angle decreases as a function of wavelength, such that $\theta_{ref} = 0°$ at $\lambda = 650$ nm, and $\theta_{ref}$ becomes increasingly negative towards $\lambda = 740$ nm, where an asymptotic branching and sign change in $\theta_{ref}$ ocurrs. Simply inverting Snell's law for wavelengths in the asymptotic region of the negative refraction band predicts effective indices in the range $\approx \pm 2$ to 3.3. Such an observation alone does not explicitly verify the sign of the refractive index, but merely the sign of energy refraction at the exit interface of the prism.

To further verify the sign of the effective index $n_{eff}$, we have retrieved the effective parameters of the six-period (twelve layer) planar metamaterial structure in Fig. 1b. The instantaneous electric and magnetic field distributions are recorded above and below the structure, and the complex reflection and transmission coefficients obtained by taking the ratios $r = E_r / E_0$ and $t = E_t / E_t$ where $E_0$ is the electric field amplitude of the incident wave and $E_r$ and $E_t$ are the amplitudes of the reflected and transmitted waves respectively. The equations for the transmission and reflection of a planar layer may be reciprocated to recover the effective impedance $z_{eff}$ (Equation 1) and index of refraction $n_{eff}$ (Equations 2 and 3) [14]. Here, $l$ is the sample thickness, $\omega$ is the angular frequency of the illuminating radiation, $c$ is the speed of light in vacuum and $m$ is the order parameter. The effective permittivity $\varepsilon$ and permeability $\mu$ are related to n and z via the relations $\varepsilon_{eff} = n_{eff} / z_{eff}$ and $\mu_{eff} = n_{eff} \cdot z_{eff}$.

$$z_{eff} = \sqrt{\frac{1 + 2r + r^2 - t^2}{1 - 2r + r^2 - t^2}} \tag{1}$$

$$n_{eff} = \frac{c}{\omega l} \sin^{-1}\left[\frac{2i}{\left(z - \frac{1}{z}\right)} \frac{r}{t}\right] + 2m\pi \tag{2}$$

$$n_{eff} = -\frac{c}{\omega l} \sin^{-1}\left[\frac{2i}{\left(z - \frac{1}{z}\right)} \frac{r}{t}\right] + (2m+1)\pi \tag{3}$$

Fig. 4 shows the real and imaginary components of the effective refractive index $n_{eff}$ as a function of wavelength. There lineshape of the refraction curve is well replicated, with agreement between the spectral positions of the negatve refraction and negative index bands. As predicted by the refraction simulation, Re($n_{eff}$) is initially positive at $\lambda = 450$ nm, crosses zero at $\lambda$ = 650 nm and is negative towards an asymptotic limit at $\lambda$ = 740 nm where a sign change occurs. The inset shows the calculated figure of merit (defined as the ratio Re($n_{eff}$)/Im($n_{eff}$)) for wavelengths in the range $\lambda$ = 650 nm – 800 nm. The maximum figure of merit exceeds 1 and is comparable to previous studies using Ag fishnet metamaterials in this wavelength range [3, 5, 15]. It should be noted that whilst large negative indices such as Re(n) = -2.7 are accompanied by a figure of merits of >2.5 they remain somewhat inaccessible due to low transmissivity of the metamaterial at these wavelengths. However, values up to Re(n) =-1 could be experimentally realised. For example, at $\lambda$ = 706 nm, the 240 nm thick metamaterial exhibits Re(n) = -0.7 and figure of merit 1.2, with the prism and planar geometries transmitting 13% and 8% respectively. Furthermore, one could also envisage further enhancing overall transmittance by optimising the metamaterial geometry to improve impedance matching [16].

An insightful decomposition of the electromagnetic response can be obtained from Figs. 4b,c where we plot the real and imaginary parts of $\varepsilon_{eff}$ and $\mu_{eff}$ as a function of wavelength. The Re($\varepsilon_{eff}$) response follows a Drude-type dispersion in which the effective plasma frequency is red-shifted compared to the bulk plasma frequency due to the dilution of the free electron concentration. The Re($\mu_{eff}$) response exhibits a single feature around $\lambda \approx$ 700 nm arising from the coupling of the incident magnetic field to the asymmetric LSPR. Away from the magnetic resonance, Re($\mu_{eff}$) tends to unity as expected for a non-resonant, non-magnetic material. The strength of the magnetic resonance is not sufficient to yield negative values of Re($\mu_{eff}$). However, as others have shown [17], Re($n_{eff}$) can be negative in a material for which real parts of Re($\varepsilon_{eff}$) and Re($\mu_{eff}$) are not simultaneously negative, provided that the condition Re($\mu_{eff}$)Im($\varepsilon_{eff}$) + Im($\mu_{eff}$)Re($\varepsilon_{eff}$) < 0 is satisfied. Changing the structural parameters of the stack (i.e. larger holes, larger pitch or reduced separation between layers) and/or introducing a high index dielectric, will cause the frequency of the asymmetric LSPR and thus the region of negative–index to shift further into the infrared.

In conclusion, we have proposed a copper metamaterial design which exhibits a negative index of refraction at visible / near-infrared wavelengths. The calculated figure of merit exceeds unity and is competitive when compared with similar Ag metamaterials operating at this wavelength range. These findings have relevance for the future design of low cost plasmonic devices and negative-index metamaterials in the visible / near-infrared.

This work is part of the research program of FOM, which is financially supported by NWO. It is also supported by the European Research Council.

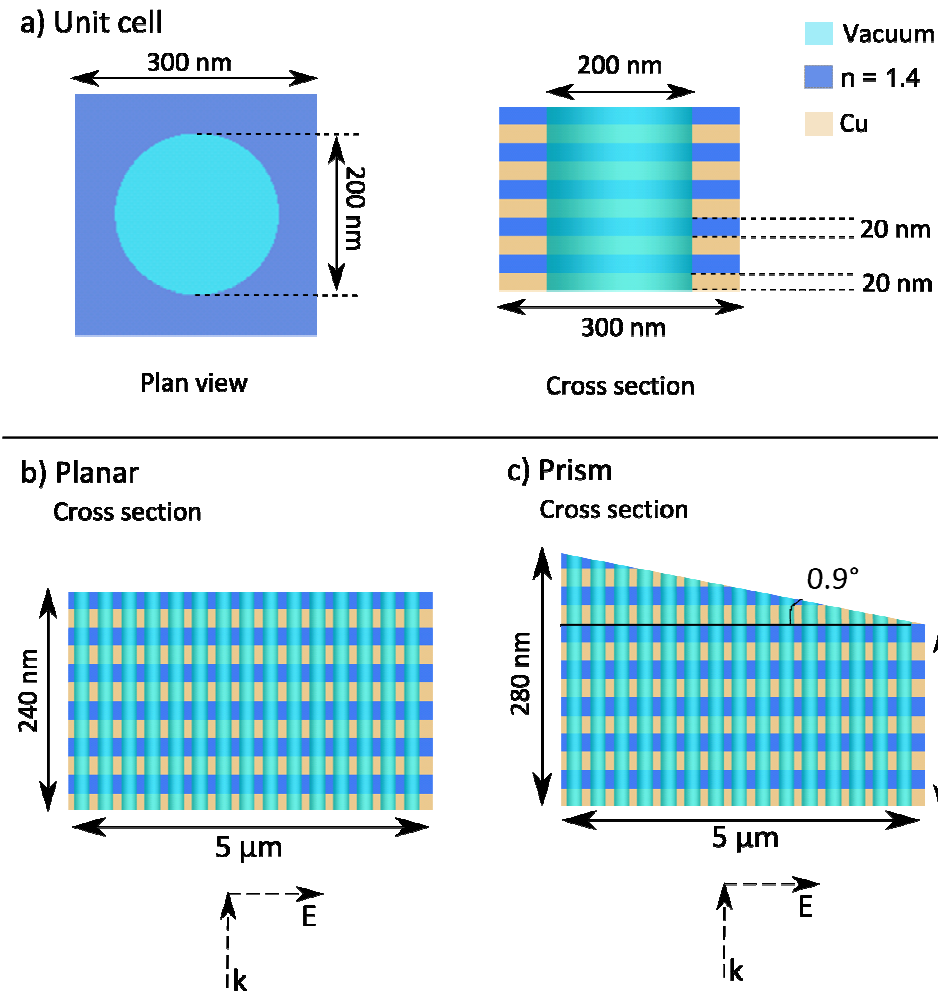

**Figure 1**. Schematic of copper metamaterial comprised of six unit-cells of 20 nm Cu / 20 nm dielectric (*n* = 1.4 + 0i). The multilayer stack is perforated with a square array of cylindrical holes (array period 300 nm, hole diameter 200 nm) which are filled with vacuum (**a,b**). To study refraction by the metamaterial, a prism is simulated in vacuum (prism angle 0.9°, Fig. 1c). The prism has a 5 μm x 5 μm square base and is tapered in thickness from seven (280 nm) to five (200 nm) unit-cells (**c**).

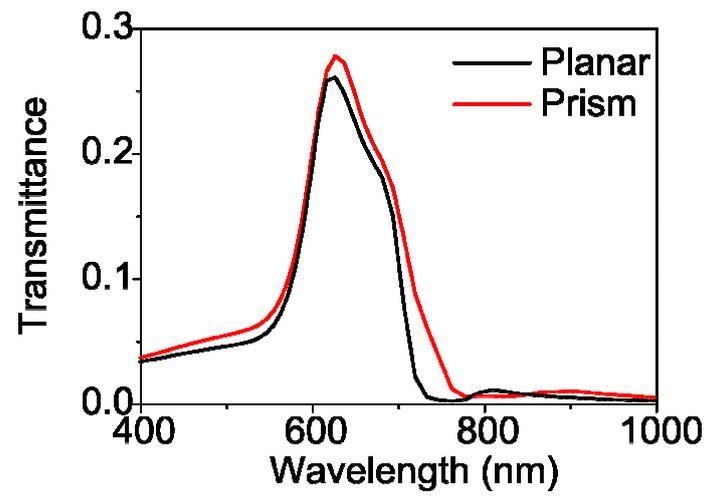

**Figure 2**. The simulated transmittance of the planar (black) and prism (red) metamaterials in Figs. 1b,c as a function of incident wavelength $\lambda$.

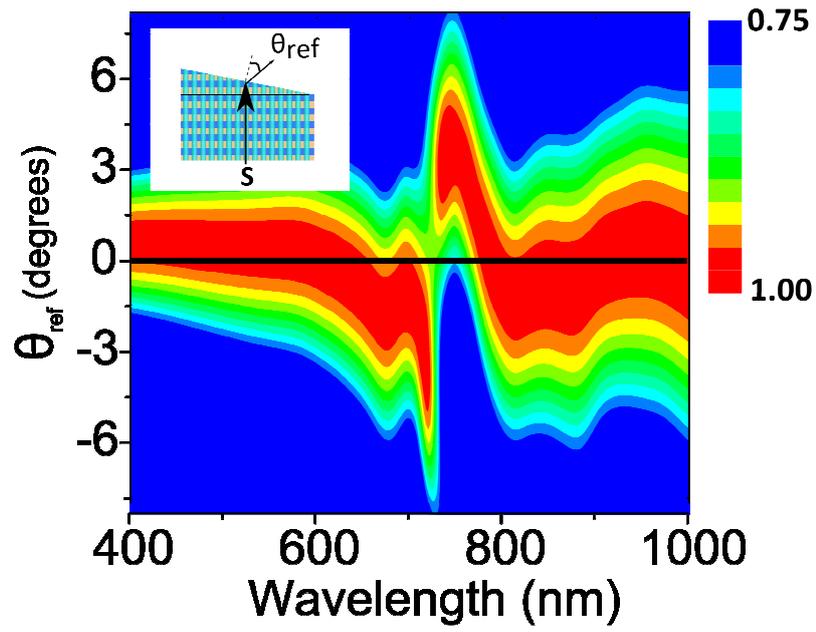

**Figure 3.** The simulated normalised electric field intensity in the far-field is plotted as a function of incident wavelength $\lambda$ and refraction angle $\theta_{ref}$ with respect to the surface normal of the prism.

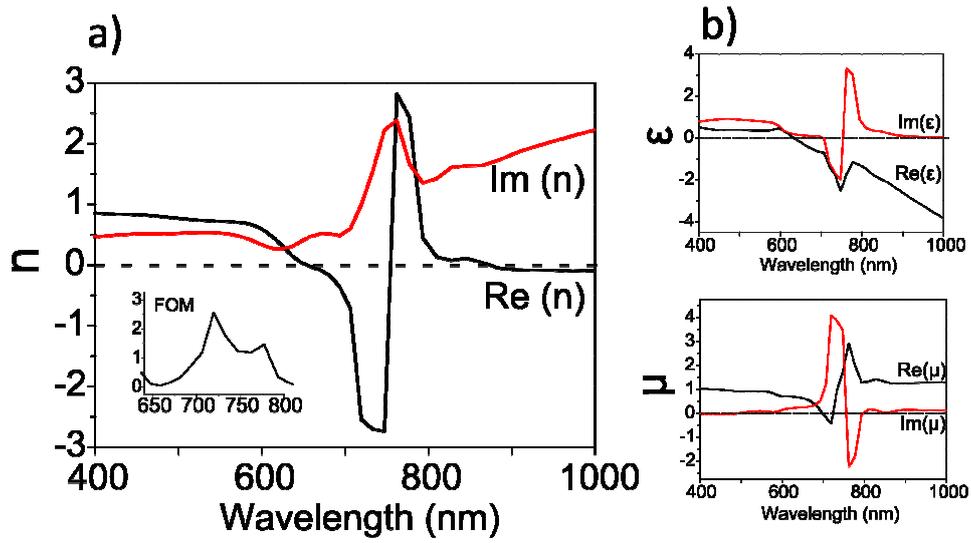

**Figure 4**. Calculated real and imaginary components of the effective refractive index $n_{eff}$, inset figure of merit (**a**) effective permittivity $\varepsilon_{eff}$ and effective permeability $\mu_{eff}$ (**b**), are plotted as a function of incident wavelength $\lambda$ for a planar six-period metamaterial.